# Query Expansion System for the VoxCeleb Speaker Recognition Challenge 2020

*Yu-Sen Cheng, Chun-Liang Shih, Tien-Hong Lo, Wen-Ting Tseng, Berlin Chen*

National Taiwan Normal University, Taiwan

{60847058S, 60847027S, teinhonglo,60847014S, berlin}@ntnu.edu.tw

## Abstract

In this report, we describe our submission to the VoxCeleb Speaker Recognition Challenge (VoxSRC) 2020. Two approaches are adopted. One is to apply query expansion on speaker verification, which shows significant progress compared to baseline in the study. Another is to use Kaldi extract x-vector and to combine its Probabilistic Linear Discriminant Analysis (PLDA) score with ResNet score.

## 1. Introduction

National Taiwan Normal University (NTNU) team's approaches to the VoxCeleb Speaker Recognition Challenge (VoxSRC) 2020 is introduced in this report. The participated track is track 1: verification fully supervised (closed). This task requires that participants train only on the VoxCeleb2 dev dataset.

Two approaches are adopted in the report. One is x-vector combined ResNet, and the other is Query Expansion (QE) [1], which is widely applied in the field of information retrieval. The x-vector is extracted by a well-known x-vector extraction topology [2]. The QE is based on the result of organizer's baseline pre-trained model. Progress in EER and DCF are obtained with both approaches compared to the baseline. Moreover, significant progress is found with QE. The results will be shown in section 4.

## 2. Experimental Setup

### 2.1. Dataset

Both x-vector and ResNet are trained on the development set of VoxCeleb2 [3], which contains 5,994 speakers. The VoxCeleb1 test sets [4], which contains 40 speakers, is the validation sets.

### 2.2. X-vector

According to previous studies [5], excellent performance on speaker verification can be obtained with Probabilistic Linear Discriminant Analysis (PLDA) [8]. Therefore, we use PLDA instead of cosine distance in the report. The model extracts the x-vector and calculates PLDA of the corresponding pair which is built on the Kaldi toolkit [2].

### 2.3. ResNet34

To save computing resources, the ResNetSE34L pre-trained model [9] is applied to extract speaker vector.

### 2.4. Rocchio algorithm

The Rocchio algorithm [10] is a query expansion method widely used in the field of information retrieval. This algorithm is applied in the study for speaker verification tasks to improve recognition performance as Eq. (1).

$$\vec{q}_m = \alpha * \vec{q} + \frac{\beta}{|D_r|} * \sum_{\forall d_j \in D_r} D_i - \frac{r}{|D_n|} * \sum_{\forall d_j \in D_n} d_j \quad (1)$$

Where $D_r$ are known to be relevant to query $\vec{q}$. On the contrary, $D_n$ are non-relevant to query $\vec{q}$. And $\alpha, \beta, \gamma$ are constants.

### 2.5. Hybrid ResNet and x-vector approach

ResNet score $S_{ResNet}$ and x-vector score $S_{Xvector}$ are combined to obtained better performance as Eq. (2).

$$S_{ResNet-Xvector} = \lambda * S_{ResNet} + (1-\lambda) * S_{Xvector} \quad (2)$$

Where $\lambda$ is introduced to specify the relative importance of ResNet score and x-vector score.

## 3. Query Expansion System

The query expansion system can be introduced with 5 parts: pre-trained model all pair scores, sort, query expansion, bidirectional calculation and evaluation.

### 3.1. Pre-trained model all pair scores

Using the pre-trained ResNet, all pair scores between all audio files can be calculated.

### 3.2. Sort

Each audio file sorts all non-self-audio file scores.

### 3.3. Query expansion

The audio files with top $n$ scores are regarded as the audio file from the same speaker, while the rest are regarded as the audio files from different speakers. We substitute Eq. (1) to calculate the speaker's centroid vector, and use this vector to recalculate the score.

### 3.4. Bidirectional calculation

In order to pursue high performance, QE is applied on both audio files.

### 3.5. Evaluation

Based on competition regulations, we use two evaluation metrics:

(i) Equal Error Rate (EER): the rate at which both acceptance and rejection errors are equal.

Table 1: The *Query Expansion result.*

| | α | β | γ | Top n | EER | DCF |
|---|---|---|---|---|---|---|
| Baseline | 1.0 | 0.0 | 0.0 | 0 | 2.365 | 0.186 |
| Query Expansion | 1.0 | 0.75 | 0.15 | 3 | 1.967 | 0.159 |
| | 1.0 | 0.75 | 0.0 | 3 | 1.988 | 0.159 |
| | 1.0 | 0.75 | 0.0 | 5 | 1.856 | 0.146 |
| | 1.0 | 0.75 | 0.0 | 10 | 1.729 | 0.132 |
| | 1.0 | 0.75 | 0.0 | 30 | 1.543 | 0.114 |
| | 1.0 | 0.75 | 0.0 | 40 | 1.522 | 0.110 |
| | 1.0 | 0.75 | 0.0 | 50 | 1.490 | 0.106 |
| | 1.0 | 0.75 | 0.0 | 60 | 1.479 | 0.107 |
| | 1.0 | 1.0 | 0.0 | 50 | 1.373 | 0.097 |
| | 0.5 | 1.0 | 0.0 | 50 | 1.235 | 0.085 |
| Bidirectional QE | 0.0 | 1.0 | 0.0 | 50 | **0.583** | **0.024** |

(ii) Minimum Detection Cost Function (MDCF): the function used by the NIST SRE [11] and the VoxSRC1 evaluations. In this competition the parameters $C_{miss} = 1$, $C_{fa} = 1$ and $p_{target} = 0.05$ are used for the cost function.

## 4. Results

In this section, we show the results of our experiment. All the inferences are tested on VoxCeleb1 test sets.

### 4.1. Query expansion

We treat the same speaker's audios as related documents. Equation (1) is applied to do query expansion. The fine-tuned parameter and corresponding results are reported in Table 1. From the result, it can be seen that EER achieve 0.584 and DCF achieve 0.024 with Bidirectional QE. The result is much better than baseline EER and DCF which is 2.365 and 0.186, respsctively.

### 4.2. x-vector

The ResNet score and x-vector score are combined with a logarithmic linear combination of Eq. (2). The results are shown in Table 2. It can be observed that EER and DCF each arrives extrema when $\lambda$ is 0.5 and 0.6, respectively.

Table 2: *x-vector combined with ResNet.*

| ResNet/X-Vector $\lambda$ | EER | DCF |
|---|---|---|
| 1.0 | 2.365 | 0.176 |
| 0.8 | 2.259 | 0.172 |
| 0.7 | 2.228 | 0.165 |
| 0.6 | 2.216 | **0.161** |
| 0.5 | **2.179** | 0.163 |
| 0.4 | 2.223 | 0.165 |
| 0.2 | 2.501 | 0.178 |
| 0.0 | 3.388 | 0.231 |

## 5. Conclusion

In this report, it is seen that QE approach shows outstanding performance on the speaker verification compared to baseline. Moreover, x-vector combined ResNet also shows progress on the speaker verification. In future work, we plan to experiment with QE on a strong baseline such as ResNetSE34V2 [9] which is a pre-trained model trained with data augmentation.